\documentclass[doublecol]{epl2} % for 2 columns style with line numbers
\usepackage{amsmath,amssymb}

\title{Transition from a dissipative to a quasi-elastic system of particles with tunable
repulsive interactions}
\shorttitle{Transition from a dissipative to a quasi-elastic system of  repulsive particles} %Insert here a short version of the title if it exceeds 70 characters

\author{S. Merminod \and M. Berhanu \and E. Falcon}
\shortauthor{S. Merminod \etal}

\institute{                    
Universit\'e Paris Diderot, Sorbonne Paris Cit\'e, MSC, CNRS (UMR 7057), 75013 Paris, France
}

\abstract{
A two-dimensional system of particles with tunable repulsive interactions is experimentally investigated. Soft ferromagnetic particles are placed on a vibrating rough plate and vertically confined, so that they perform a horizontal Brownian motion in a cell. When immersed in an external vertical magnetic field, the particles become magnetised and thus interact according to a dipolar repulsive law. As the amplitude of the magnetic field is increased, magnetic repulsion raises and the rate of inelastic collisions decreases. Studying the pair correlation function and the particle velocity distributions, we show that the typical properties of such a dissipative out-of-equilibrium granular gas are progressively lost, to approach those expected for a usual gas at thermodynamic equilibrium. For stronger interaction strengths, the system gradually solidifies towards a hexagonal crystal. This new setup could consequently be used as a model experimental system for out-of-equilibrium statistical physics, in which the distance to the quasi-elastic limit can be accurately controlled.}

\begin{document}

\maketitle

\section{Introduction}

Statistical mechanics provides, with the assumption of thermodynamic equilibrium, a precise description of molecular gases composed of thermally agitated microscopic particles. In contrast, in granular gases macroscopic particles are mechanically driven. Since the collisions between these particles are dissipative, energy must be continuously provided into the system from outside to reach a stationary out-of-equilibrium state. In consequence, granular gases have been extensively studied as a model system for out-of-equilibrium statistical physics~\cite{Barrat2005} theoretically~\cite{NoijeErnst1998,Goldhirsch1993}, numerically~\cite{Moon2001} and experimentally~\cite{Olafsen1998,Olafsen1999,Losert1999}. Two-dimensional granular gases, \textit{i.e.}, particles lying on a horizontal plate vertically vibrated, were especially studied because particle trajectories can be reconstructed using fast imaging and tracking algorithms ~\cite{Olafsen1998,Losert1999,Olafsen1999,Reis2007,Puglisi2012}. Nevertheless, few studies investigated the case in which non-contact interactions between particles compete with kinetic agitation and thus introduce spatial correlations differing from those observed for an inelastic hard-sphere gas. In a granular gas composed of particles owing a permanent magnetic dipole, the anisotropic dipole-dipole interactions lead to dipole alignment, then attraction and clustering~\cite{Blair2003,Stambaugh2003,Oyarte2013}. In contrast, physics differs strongly using ferromagnetic particles with a low remnant magnetic field. When immersed in an external static magnetic field, such particles acquire an induced magnetisation so that inter-particle dipolar interactions become tunable by the operator. Applying this protocol to a granular packing, a first order fluid-solid transition~\cite{Laroche2010} and a surface instability due to competition between gravity and magnetic forces~\cite{Lopez2010}, are observed. If such particles are confined in a two-dimensional plane and immersed in an external perpendicular magnetic field, the magnetic interactions between particles are purely repulsive, since their dipoles are all aligned in the vertical field direction.  At low packing fraction, low agitation and high magnetic field, the system forms a hexagonal lattice~\cite{Schockmel2013}. As mechanical agitation is increased crystal melting is observed, that is, translational and orientational orders disappear, as in some other 2D systems of interacting particles~\cite{Ghazali2006,Dillmann2012,Deutschlander2013,Boyer2009,Coupier2005}.\\
\indent In this letter, we study a 2D granular gas with such tunable repulsive magnetic interactions. To our knowledge, the influence of dipolar interactions on the particle velocity distributions has only been studied in a case dominated by attractive interactions~\cite{Kohlstedt2005}. Free-cooling of 3D granular gases with electrostatic repulsions has also been investigated theoretically and numerically~\cite{SchefflerWolf2002,MuellerLuding2011}. In our experiment, we start from the well studied case of a two-dimensional granular gas~\cite{Reis2007,Puglisi2012} where mechanical agitation is provided to the particles by the vibration of a horizontal rough bottom plate. 
Additionally, a vertical magnetic field is then applied, leading to repulsive dipolar interactions between particles. Using particle tracking techniques, we analyse quantitatively the structural changes within the granular gas and its dynamical properties. The rate of inelastic collisions between particles can be easily tuned. Indeed, increasing the amplitude of the magnetic field enhances magnetic repulsion and thus decreases the number of inelastic collisions. As the dissipation rate due to inelastic collisions is proportional to the number of collisions, the total dissipation in the system is reduced. We thus show that the system undergoes a transition from a dissipative to a quasi-elastic system when the magnetic field is increased.
\begin{figure}[t!]
	\begin{center}
	\includegraphics[width=8.6cm]{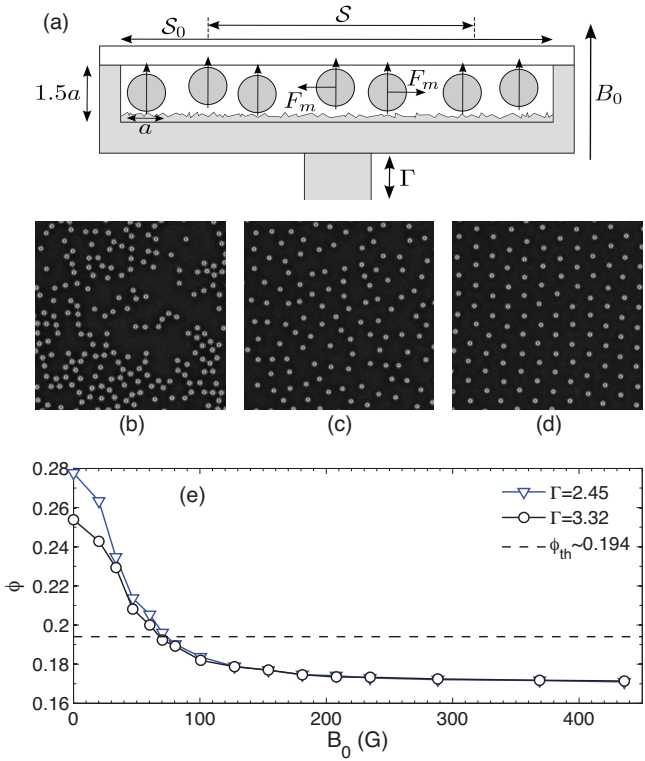}
    \end{center}
    	\caption{(color online) (a)~Experimental setup (see text).  (b)~Snapshots in the inelastic regime for $B_0=0\,$G ($\varepsilon=0$), (c)~the quasi-elastic regime for $B_0=127\,$G ($\varepsilon=16.2$), and (d)~the hexagonal crystal-like regime for $B_0=436$\,G ($\varepsilon=283$). Snapshots size is $3.5$\,cm\,$\times$\,$3.5$\,cm, and $\Gamma=3.32$. For the full time evolution, see \texttt{movie1.m4v} in supplementary material. (e)~Area fraction of balls $\phi$ in the area $\mathcal{S}$ versus $B_0$ for $\Gamma=2.45$ and $3.32$. The dashed line corresponds to the area fraction for a homogeneous particle distribution $ \phi_{\scriptsize \textrm{th}}=N_0 \pi \sigma^2 / \mathcal{S}_0 \approx 0.194 $.}
	\label{schemab}
	\label{phiVSB}
	\label{Photo1}
	\label{Photo2}
\end{figure} 

\section{Experimental setup}
The experimental cell is depicted in Fig.~\ref{schemab}(a). It consists of a horizontal, square duraluminium bottom plate of area $\mathcal{S}_{0}=9\,\textrm{cm}\times9\,$cm and covered by a sandpaper sheet in order to provide roughness (RMS amplitude of $20\,\mu$m). The cell is filled with $N_0 = 2000$ chrome steel (AISI~52100) spherical particles with a diameter $a = 2\,\sigma = 1\,\textrm{mm}\pm 2.5\,\mu$m and a mass $m=4.07\times10^{-6}$\,kg. These balls are confined by rigid aluminium walls and by a rigid, smooth, antistatic coated polycarbonate lid placed $1.5a$ above the bottom plate. In order to reach a non-equilibrium steady state, this cell is driven sinusoidally in the vertical direction by means of an electromagnetic shaker. The dimensionless acceleration is $\Gamma \equiv (2\pi f)^2 A/g $ with $f = 300\,$Hz the frequency and $A$ the amplitude of the sinusoidal forcing, $g$ being the gravitational acceleration. $\Gamma$ is measured  using an accelerometer screwed on the cell. Two coils generate a vertical magnetic field $\boldsymbol{B_0}$ which is perpendicular to the cell plane and is spatially homogeneous within the cell volume with a $2$\,\% accuracy. A high speed camera (Phantom V10) is located above the centre of the cell. A diffusive LED ring encircling the cell illuminates from the top the particles that appear as bright rings on a dark background. The camera acquisition rate is fixed to $779$~frames per second in order to detect the collisions between particles. Video recordings are performed once the stationary state is reached (waiting time of 60~s) and last at least 3.85~s. To avoid measurement issues at the boundaries, we choose a region of interest $\mathcal{S}$ of 5.7$\,$cm$\times\,$5.7$\,$cm around the cell centre. The particle diameter then corresponds to 20~pixels. We performed individual detection of particles from the video recordings using first a convolution-based least-squares fitting particle detection routine~\cite{Shattuck,Reis2007} completed by an intensity-weighted centre detection algorithm. This provides particle centre positions with a resolution of less than 0.3$\,$pixel$\,\sim\,$0.015$a$ \cite{Puglisi2012}. Finally, individual trajectories were reconstructed using a tracking algorithm~\cite{CrockerWeeks,BlairDufresne}. Hence, from highly resolved particle position data, we compute their velocity distributions, pair correlation functions, mean square displacements as well as collision rate estimations.

\section{Experimental parameters}
Let us now describe the influence of the external magnetic field $\boldsymbol{B_0}$ on the chrome steel particles. These balls are soft ferromagnetic, \textit{i.e.}, with a low remnant magnetic field and a high magnetic permeability. When placed in a vertical magnetic field of amplitude $B_0$, each particle is uniformly magnetised. It behaves as an induced magnetic dipole of magnetic moment $\frac{4}{3}\pi \sigma^3 \frac{\chi_m}{\mu_0} B_0\,\boldsymbol{e_z}$, with $\chi_m$ the volume magnetic susceptibility, $\mu_0$ the vacuum permeability, and $\boldsymbol{e_z}$ the upward unit vector along the vertical axis. For a purely 2D system of two identical spheres $i$ and $j$ with  $\boldsymbol{B_0}$ perpendicular to $\boldsymbol{r_{ij}}$ (the horizontal vector between the particle centres), the potential energy of magnetic interaction reads~\cite{Jackson}:
\begin{equation} 
E_{m,\,\left\langle i,j\right\rangle} = \frac{4 \pi}{\mu_0}\, {B_0}^2 \, \frac{\sigma^6}{\vert \boldsymbol{r_{ij}} \vert ^3}
\label{eq:Em}
\end{equation}
in the limit of high intrinsic magnetic permeability. We point out that without taking into account the geometry of the magnetisation and the demagnetising magnetic field, an effective susceptibility $\chi$ can be defined~\cite{Schockmel2013,Deutschlander2013}, yielding an expression of the magnetic energy proportional to Eq.~(\ref{eq:Em}). The repulsive force between these two particles, $\boldsymbol{F_{m,\,\left\langle i,j\right\rangle}}= - \boldsymbol{\nabla} \, E_{m,\,\left\langle i,j\right\rangle}$,
decreases with ${\vert \boldsymbol{r_{ij}}\vert}^{-4}$ and is directed along $\boldsymbol{r_{ij}}$. Therefore the repulsion between particles can be tuned by the amplitude of the magnetic field $B_0$. If particles are not exactly in the same horizontal plane between the two confining plates, the horizontal repulsive force is reduced due to 3D effects. Nevertheless, further results in this letter show that a 2D analysis is relevant to describe the system behaviour, by considering $ E_{m,\,\left\langle i,j\right\rangle} $ from Eq.~(\ref{eq:Em}) as a scale of the actual magnetic energy of two interacting particles.\\
\indent In addition to parameters $\Gamma$ and $B_0$, the last important parameter is the dimensionless area fraction $ \phi \equiv N\, \pi \sigma^2/\mathcal{S}$, with $N$ the number of particles detected in the region of interest $\mathcal{S}$. As it can be expected for a system of particles with increased repulsive interactions, we observe~[Fig.~\ref{schemab}(e)] an expansion of the system when $B_0$ is increased. $\phi$ is found to be a decreasing function of $B_0$, which differs from the expected value $\phi_{\scriptsize\textrm{th}} = N_0\,\pi\sigma^2 / \mathcal{S}_{0} \approx 0.194$ computed for the full cell area. Indeed, $\phi \approx 0.27$ for $B_0 \approx 0$\,G due to clustering~\cite{Goldhirsch1993,Olafsen1998,Prevost2004,Falcon1999,FalconWunenburger1999} in the cell central region. As $B_0$ is increased, the horizontal magnetic repulsive forces cause the granular gas to expand and to reach a state of smaller and homogeneous area fraction in the region of interest $\mathcal{S}$. It is well known that a higher particle density near the boundaries is induced by non-repulsive boundary conditions \cite{SchefflerWolf2002} and a weak magnetic field radial gradient. Nevertheless, we point out that $\phi$ is found to be homogeneous in the region of interest $\mathcal{S}$ whatever $B_0 > 0$, the inhomogeneity of $\phi$ being confined within the area outside~$\mathcal{S}$.
\begin{figure}[t!]
	\begin{center}
	\includegraphics[width=8.6cm]{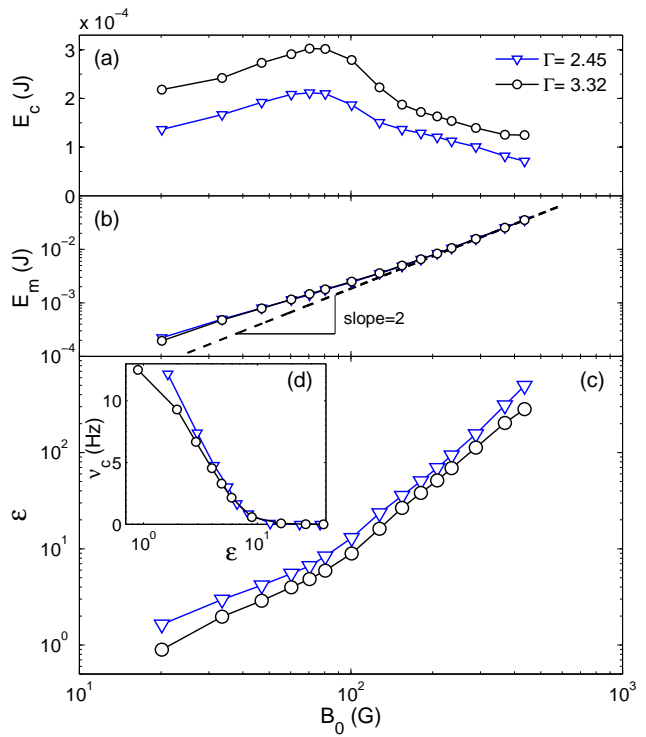}
	\end{center}
    \caption{(color online) (a)~Particle kinetic energy $E_c$  as a function of the magnetic field $B_0$ for accelerations $\Gamma = 2.45$ and $3.32$. (b)~Particle magnetic potential energy $E_m$ versus $B_0$. (c)~Ratio of the energies $\varepsilon= E_m/E_c$ versus $B_0$. (d)~Collision rate $\nu_c$ (number of collision per particle and time unit) versus~$\varepsilon$.}	
    \label{fig:em_ec_eps}
    \label{energy}        	
    \label{fig:tauxColl}    
\end{figure}

\section{Competition between kinetic and magnetic energies}
From the parameters $B_0$, $\Gamma$ and $\phi$, we define now the relevant physical quantities, that we use to describe the behaviour of our system. Considering a 2D assembly of $N$ particles mechanically agitated and immersed in $B_0$ inside the region of interest $\mathcal{S}$, we compute its kinetic energy per particle from velocity measurements, namely $E_c=\frac{1}{2} m \, \overline{\langle {v_x}^2+{v_y}^2 \rangle}$, where $v_x$ (resp. $v_y$) denotes the horizontal velocities in the $x$-direction ($y$-direction), $\langle\cdot\rangle$ an ensemble average and $\overline{\;\cdot\;\rule[0mm]{0mm}{2mm}}$ the temporal average. Note that $E_c$ is directly proportional to the granular temperature usually defined as $T_g = \frac{E_c}{m}$~\cite{Reis2007,Puglisi2012}. We also compute the magnetic energy per particle $E_{m} = \overline{\frac{1}{N} \, \sum_{i=1}^{N}\sum_{j=i+1}^{N} E_{m, \,\left\langle i,j \right\rangle}}$, with $\left\langle i,j \right\rangle$ a pair of particles within $\mathcal{S}$ and $E_{m, \,\left\langle i,j \right\rangle}$ its potential energy from Eq.~(\ref{eq:Em}). The magnetic potential energy depends on the local configuration of the particles, and therefore it fluctuates in time. Finally, a dimensionless interaction parameter is defined by the ratio $\varepsilon\equiv E_m/E_c$ between the magnetic and kinetic energies~\cite{Deutschlander2013,Schockmel2013}. When $\varepsilon$ is increased, the system undergoes a continuous transition from an inelastic granular gas [Fig.~\ref{Photo1}(b)] to a quasi-elastic granular gas [Fig.~\ref{Photo1}(c)] since inelastic collisions between particles are progressively replaced by elastic magnetic interactions. At higher $\varepsilon$, the system self-organises in a condensed-like phase showing a 2D-hexagonal crystal lattice [Fig.~\ref{Photo2}(d)] as particle displacements become constrained due to magnetic repulsions. This evolution of the system is also shown in the supplementary material \texttt{movie1.m4v} for a continuous increase of $B_0$ at fixed $\Gamma$.\\ 
\begin{figure}[t!]
 	\begin{center}
 	\includegraphics[width=8.6cm]{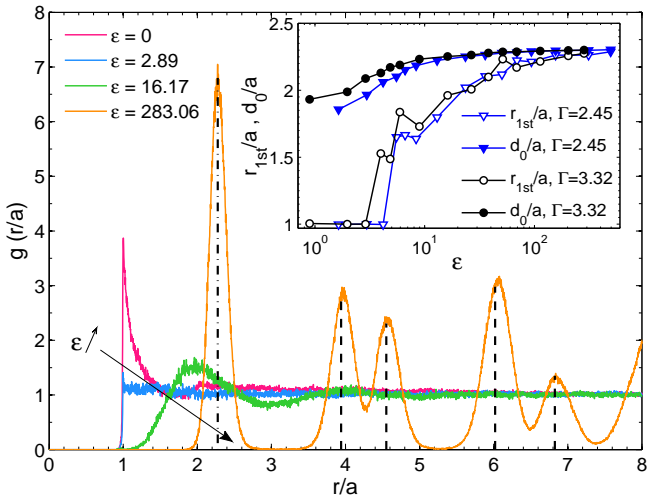} 
    	\caption{(color online) Pair correlation function $g(r)$ for $\Gamma=3.32$ and $\varepsilon=0$, $2.89$, $16.2$ and $283$. For the last value, peak positions for a hexagonal crystal are shown in dashed lines as multiples of the lattice distance  $r_{\scriptsize \textrm{1st}}/a=2.28$, for factors $1, \sqrt{3}, 2, \sqrt{7}$, and $3$. Inset: Position of the first peak $r_{\scriptsize \textrm{1st}}/a$ versus $\varepsilon$ and compared with $d_0/a$ expected for a hexagonal crystal (see text).}
    \label{paircorell}
       \end{center}
 \end{figure}
\indent We present now experimental results obtained for increasing $B_0$ and for fixed $\Gamma$ ($2.45$ or $3.32$). These values correspond to the bounds of the range of $\Gamma$ where $E_c$ increases linearly~\cite{Reis2007}. The evolutions of $E_c$, $E_m$ and $\varepsilon$ with $B_0$ are depicted in Fig.~\ref{energy}(a)-\ref{energy}(c). Note that $\varepsilon$ is larger than 1 for $B_0 > 20$\,G, meaning that regimes dominated by magnetic repulsions are reached for moderate values of $B_0$. We also notice a non-monotonous evolution of $E_c$, which reaches a maximum for $B_0 \approx 70$\,G ($\varepsilon \approx 5$). The rate of inelastic collisions between particles $\nu_c$~is indeed strongly reduced for increasing values of $\varepsilon$ as depicted in Fig.~\ref{fig:tauxColl}(d). Due to the magnetic energy barrier, only particles with sufficient kinetic energy can collide~\cite{SchefflerWolf2002}. The average number of collisions per particle and per time unit, $\nu_c$, is evaluated using an algorithm detecting individual collisions through a distance criterion selective process. $\nu_c$~decreases with $\varepsilon$, vanishes below $0.1$~Hz for $\varepsilon > 10$ and is strictly zero for $\varepsilon > 30$. For greater $\varepsilon$, particle displacements become bounded~\cite{Schockmel2013}. Indeed, in Fig.~\ref{energy}(b) $E_m$ is found to be proportional to ${B_0}^2$ for $B_0 > 150$\,G ($\varepsilon > 30$), because particle geometrical arrangement becomes a fixed parameter in the calculation of $E_m$, once they are magnetically confined.

\section{Radial pair correlation function}
This energetic description is also associated with important structural changes, which can be enlightened by computing the radial pair correlation function $g(r)~\equiv~\overline{ \left[ \sum_{i=1}^{N}\sum_{j\neq i} \delta (r - r_{ij}) \right] \mathcal{S}/(2\pi r N^2) }$, with $r_{ij}$ is the distance between the particles $i$ and $j$. This function gives the probability to find two particle centres separated by a distance $r$. $g(r)$ is shown in Fig.~\ref{paircorell} for characteristic values of $\varepsilon$ and at fixed $\Gamma$. At $\varepsilon=0$, $g(r)$ displays a sharp peak at the contact value $r=a$, as in usual granular gases~\cite{Olafsen1998,Olafsen1999}. This confirms that most collisions occur in horizontal planes and validates the 2D description. Collisions happening out of horizontal planes, when viewed from the top, produce indeed a partial overlapping, leading to non-vanishing values of $g(r)$ for $r<a$. When $\varepsilon$ is slightly increased, the amplitude of the first peak decreases to almost 1, giving a nearly flat $g(r)$ (see the curve for $\varepsilon$~=~2.89). This shows that radial correlations are then quasi-absent as for a non-dissipative perfect gas whose $g(r)$ is zero for $r<a$ and~1 elsewhere in the vanishing density limit. When $\varepsilon$ is further increased, this feature is gradually lost. Due to magnetic repulsions, $g(r=a)$ decreases towards zero and a first peak appears at $r>a$, indicating the appearance of a preferential distance between particles. A similar transition of $g(r)$ has been observed numerically for a 3D repulsive granular gas with a Coulomb interaction potential~\cite{SchefflerWolf2002}. For high enough values of $\varepsilon$, the system structure approaches the one of a hexagonal crystal~\cite{Schockmel2013}. In this case, once the lattice cell size is set to the first peak position, theoretical secondary peak positions can be predicted from geometrical calculations and are indeed found to be close to the measured values (see the vertical dashed lines in~Fig.~\ref{paircorell}). The dimensionless position of the first peak of the pair correlation function $r_{\scriptsize \textrm{1st}}/a$ versus $\varepsilon$ [Fig.~\ref{paircorell}(inset)] can be used to discriminate the different regimes. Indeed, for $\varepsilon < 2.89$ ($\Gamma=3.32$), $r_{\scriptsize \textrm{1st}}/a=1$, which corresponds to a gas-like state becoming more and more elastic as $\varepsilon$ increases. Then, for higher values of $\varepsilon$, $r_{\scriptsize \textrm{1st}}/a > 1$ means that a fluid-like phase with a negligible collision rate is reached. A system solidification progressively occurs: $r_{\scriptsize \textrm{1st}}/a$ grows slowly with $\varepsilon$ and gradually approaches the value expected for the hexagonal lattice $d_0/a~=~\sqrt{\pi/(2\sqrt{3} \phi)}$, which depends on $\phi$ since measured in $\mathcal{S}$. A distance to the hexagonal crystal is thus provided by the calculation of $d_0/a - r_{\scriptsize \textrm{1st}}/a$.
\\ \indent Recently, such a crystal formation has also been observed in a 2D granular system of repulsive particles~\cite{Schockmel2013}, and this crystal was found to melt through a hexatic phase in good agreement with the Kosterlitz-Thouless-Halperin-Nelson-Young (KTHNY) scenario~\cite{Strandburg1988}. In our experiments, the computations of the pair correlation function and of the orientational correlation function (not shown here) lead to qualitatively similar results as in ~\cite{Schockmel2013}. Finally, the behavior of the collisionless nearly crystalline state at strong enough $\varepsilon$ can be understood as follows. We can consider our non-contact repulsive particles as effective larger particles in a close packing of disks. Their effective diameter would be given by $r_{\scriptsize \textrm{1st}}$, the first-peak position of $g(r)$, leading to an effective area fraction $\phi_{\scriptsize \textrm{eff}}=(r_{\scriptsize \textrm{1st}}/a)^2\,\phi$, varying roughly between $0.44$ and $0.90$. Therefore, when the system is collisionless, increasing $\varepsilon$ can be understood as rising the effective density $\phi_{\scriptsize \textrm{eff}}$. This explains why transitions similar to those of 2D close-packed particle systems~\cite{Olafsen2005} might be found in our study where non-contact interactions between particles are involved.

\section{Mean square displacements}
\begin{figure}[t!]
 	\begin{center}
 		\includegraphics[width=8.6cm]{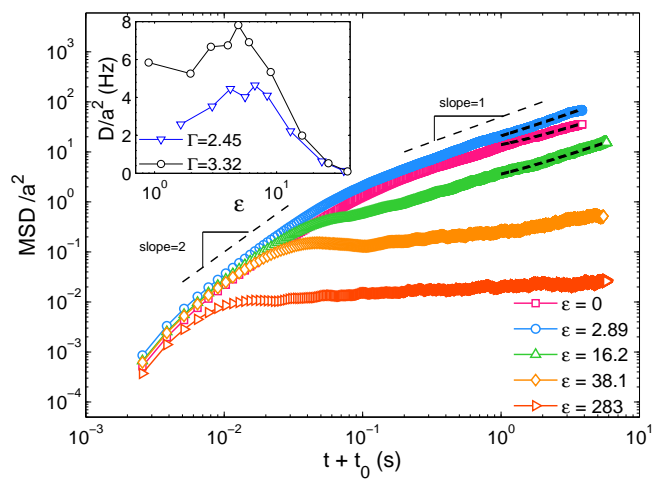} 
    		\caption{(color online) Mean square displacements (MSD) for $\Gamma=3.32$ and $\varepsilon=0$, $2.89$, $16.2$, $38.1$ and $283$. The thin dashed lines indicating slopes of values 1 and 2 are guides to the eye. The thick dashed lines superimposed on MSD data from $t=1$\,s to the end of the recordings are linear fits performed in the normal diffusive regime. Inset: slopes of the linear fits divided by 4, $D$, which can be identified with a diffusion coefficient.}
    		\label{fig:msd_d}
 	\end{center}
\end{figure}

Another way to characterise structural and dynamical changes consists in measuring the mean square displacements (MSD) of the particles $\left\langle \vert  \boldsymbol{R}(t+t_0) - \boldsymbol{R}(t_0) \vert^2 \right\rangle$, where $\boldsymbol{R}(t)$ is the particle position at time $t$, $t_0$ being an arbitrary time origin. For particles experiencing a Brownian motion in two dimensions, the MSD equals $4D_b t$, where $D_b$ is the diffusion coefficient. MSD normalised by the particle diameter are plotted in Fig.~\ref{fig:msd_d}. For $\varepsilon<30$, at short times a ballistic regime occurs (MSD $\propto t^2$), followed by a normal diffusive regime at longer times  (MSD $\propto t$). Therefore, in this regime, particles perform a horizontal quasi-Brownian motion in the experimental cell. For $\varepsilon > 30$, the diffusion becomes anomalous : a fit of the MSDs by a power law $t^\alpha$ would provide $\alpha<1$, showing that particles undergo a sub-diffusive motion. We point out that simultaneously, the collision rate becomes zero, marking a change of behaviour of the particles as magnetic interactions become stronger. Moreover, the derivative of the MSD vanishes at finite times as the MSD locally saturates, shedding light onto the existence of magnetic confinement. This becomes very clear for $\varepsilon>10^2$, as particles are strongly confined and move around equilibrium positions corresponding to the nodes of the hexagonal lattice. \\
\indent For $\varepsilon \leq 30$ and after waiting long enough to define a normal diffusive regime, we extract from the MSD the particle diffusion coefficient $D$, computed as one fourth of the slope of the MSD (evaluated from $t+t_0=1$\,s until the end of the measurement). The corresponding fits are plotted as thick dashed lines in  Fig.~\ref{fig:msd_d}, and the obtained values of $D$ are shown in Fig.~\ref{fig:msd_d}(inset). Like $E_c$ [Fig.~\ref{fig:em_ec_eps}(a)], $D$ as a function of $\varepsilon$ is non monotonous and decreases strongly for $\varepsilon \gtrsim 5$ (\textit{i.e.}, $B_0 \gtrsim 70$ \,G), showing that magnetic repulsions oppose the displacements. It can also be noticed that $D$ and $E_c$ reach their respective maximum for values of $\varepsilon$ of the same order of magnitude, when repulsive interactions are of the same order as kinetic agitation. Indeed, $D$ can be roughly evaluated as the product of the root mean square velocity, which is directly related to $E_c$, by the mean free path, which should decrease with $\phi$ and $\varepsilon$ as the magnetic confinement opposes the particle displacements. The evolutions of $D$ and $E_c$ are thus deeply~connected.

\section{Velocity statistics}
\begin{figure}[t!]
	\begin{center}
		\includegraphics[width=8.6cm]{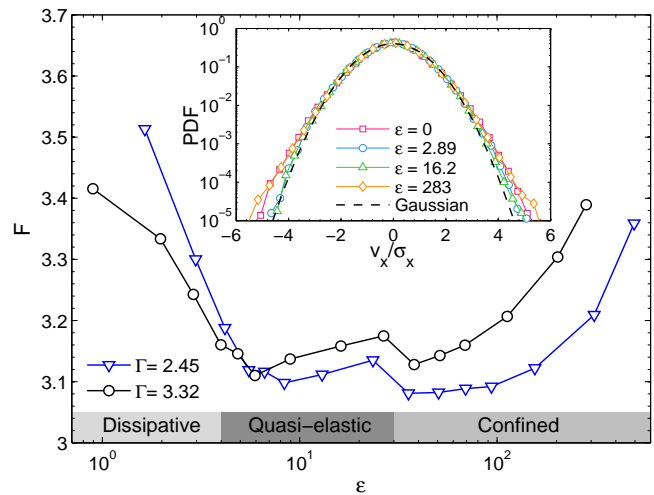}
       	\caption{(color online) Flatness of the velocity probability density functions as a function of $\varepsilon$. For $B_0=0$\,G (\textit{i.e.}, $\varepsilon\rightarrow -\infty$), $F~=~3.52$ for $\Gamma=3.32$ (resp. $F=3.63$ for $\Gamma=2.45$). Inset: velocity PDFs for different  $\varepsilon$ at $\Gamma = 3.32$. The dashed curve is the Gaussian distribution.} 
		\label{pdfv}
	\end{center}
\end{figure}

Structural modifications imply important changes on dynamics, especially on the particle trajectories and velocity distributions. In the inset of Fig.~\ref{pdfv}, the probability density functions (PDFs) of velocities ($x$-component) normalised by their standard deviation, $v_x/\sigma_x$ with $\sigma_x \equiv \sqrt{\langle {v_x}^2 \rangle}$, are plotted at fixed $\Gamma$ for different values of $\varepsilon$. They are compared to the Gaussian distribution expected for a perfect gas at thermal equilibrium. Identical results are found for $v_y$ due to system isotropy in the central region. As predicted for an infinite system~\cite{NoijeErnst1998} and reported in other experiments~\cite{Olafsen1999,Losert1999,Reis2007,Puglisi2012}, at $\varepsilon=0$ the velocity distribution presents a deviation from the Gaussian. 
In fact, this behavior is expected for out-of-equilibrium systems, as Gaussian distribution is predicted for equilibrium gases. The reported overpopulation of the high-velocity tails is characteristic of granular gases, although there is no simple argument to justify it~\cite{Barrat2005}. As $\varepsilon$ is increased, the PDFs become progressively closer to the Gaussian until $\varepsilon \approx 10$ but then depart for higher values.
\\ \indent This behaviour is better depicted  by plotting  the flatness of the velocity distributions, defined as $F \equiv \left\langle (v_x- \left\langle v_x \right\rangle)^4 \right\rangle / \sigma_x^4$ and shown in Fig.~\ref{pdfv}. For a purely Gaussian distribution $F$ equals $3$ and is larger for more spread distributions. A range of significantly low values of $F$ can be defined for $4 < \varepsilon< 30$, where the granular gas can be considered as quasi-elastic. Indeed, energy exchanges between particles should occur mainly through magnetic repulsive interactions, which are dissipationless. Note that the lower bound in $\varepsilon$ is fairly consistent with the value $\varepsilon=2.89$ separating the usual granular gas regime and the one with negligible collisions (see Fig.~\ref{paircorell}). For $\varepsilon > 30$, displacements become progressively constrained by magnetic repelling and the system can be seen as an assembly of confined particles~\cite{Schockmel2013}. $F$ then increases with $\varepsilon$, highlighting a heterogeneity of velocities, as particles are individually more or less confined.

\section{Conclusion}
We have studied the effect of tunable repulsive dipolar interactions on a quasi-two-dimensional granular gas. For fixed dimensionless accelerations $\Gamma$ and in a low density regime ($\phi \approx 0.2$), we increased the magnetic field $B_0$. The rise of the ratio $\varepsilon$ between magnetic interaction and kinetic agitation leads to a continuous phase transition from a dissipative granular gas state at $\varepsilon$~=~0, to a collisionless hexagonal nearly crystalline state at high~$\varepsilon$. More interesting, in the intermediate range of $\varepsilon$, structural and dynamical properties of the magnetic granular gas display similar features to those expected for a molecular gas at thermal equilibrium (quasi-Gaussian velocity distributions and nearly flat pair correlation functions). This transition from a dissipative to a quasi-elastic granular gas, when $B_0$ is increased, comes from the decrease of the dissipative collision rate, which leads to the reduction of the total dissipation. Hence, the 2D granular gas is then closer to the quasi-elastic limit. We were thus able to produce a macroscopic system whose distance to the quasi-elastic limit could be precisely controlled through the applied magnetic field. We may also wonder how the results found here with repulsive dipolar interactions can be generalised for other interaction potentials, like the Coulombian one~\cite{SchefflerWolf2002,MuellerLuding2011}. \\ %
\indent Future studies on this new system could be useful to validate theoretical works about out-of-equilibrium dissipative gases, by investigating velocity correlations and coupling with the forcing viewed as a thermal bath~\cite{Gradenigo2011,Puglisi2012}. Another perspective is to apply a magnetic quench to the system, in order to try to solidify it into a disordered state, which could be analogous to a colloidal glass~\cite{PuseyvanMegen1986}. Moreover, for denser regimes and for high $\varepsilon$, we observe other complex disordered states. Our experimental system could indeed be used to mimic, at the macroscopic scale, geometric frustration~\cite{HanNature2008,ShokefPRL2009} or topological defects~\cite{YaoPRL2013} arising in various physical systems.

\acknowledgments
We thank J.-C.~Bacri for providing the coils and for discussions, and P.~Visco, L.~Deike, L.~Gordillo and T.~Jamin for fruitful discussions. This work has been supported by Universit\'{e} Paris Diderot (BQR 2012, UFR Physique) and by ESA Topical Team on granular materials No.~4000103461.


\begin{thebibliography}{0}

\bibitem{Barrat2005}
  \Name{Barrat A., Trizac E., and Ernst M. H.}
  \REVIEW{J. Phys. : Condens. Matter}{17}{2005}{99}.

\bibitem{NoijeErnst1998}
  \Name{van Noije T. P. C. and Ernst M. H.}
  \REVIEW{Gran. Mat.}{1}{1998}{57}.
  
\bibitem{Goldhirsch1993}
  \Name{Goldhirsch I. and Zanetti G.}
  \REVIEW{Phys. Rev. Lett.}{70}{1993}{1619}.
  
\bibitem{Moon2001}
  \Name{Moon S. J., Shattuck M. D. and Swift J. B.}
  \REVIEW{Phys. Rev. E.}{64}{2001}{031303}.
  
\bibitem{Olafsen1998}
  \Name{Olafsen J. S. and Urbach J. S.}
  \REVIEW{Phys. Rev. Lett.}{81}{1998}{4369}.

\bibitem{Olafsen1999}
  \Name{Olafsen J. S. and Urbach J. S.}
  \REVIEW{Phys. Rev. E}{60}{1999}{R2468}.
  
\bibitem{Losert1999}
  \Name{Losert W. {\it et al.}}
  \REVIEW{Chaos}{9}{1999}{3}.
  
\bibitem{Reis2007}
  \Name{Reis P. M., Ingale R. A., and Shattuck M. D.}
  \REVIEW{Phys. Rev. E}{75}{2007}{051311}.
  
\bibitem{Puglisi2012}
  \Name{Puglisi A. {\it et al.}}
  \REVIEW{J. Chem. Phys.}{136}{2012}{014704}.


% 10th  
\bibitem{Blair2003}
  \Name{Blair D. L. and Kudrolli A.}
  \REVIEW{Phys. Rev. E}{67}{2003}{021302}.
  
\bibitem{Stambaugh2003}
  \Name{Stambaugh J., Lathrop D. P., Ott E., and Losert W.}
  \REVIEW{Phys. Rev. E}{68}{2003}{026207}.
  
\bibitem{Oyarte2013}
  \Name{Oyarte L., Guti\'{e}rrez P., Auma\^{i}tre S., and Mujica N.}
  \REVIEW{Phys. Rev. E}{87}{2013}{022204}.
  
\bibitem{Laroche2010}
  \Name{Laroche C. and P\'{e}tr\'{e}lis F.}
  \REVIEW{Eur. Phys. J. B}{77}{2010}{489}.
  
\bibitem{Lopez2010}
  \Name{Lopez D. and P\'{e}tr\'{e}lis F.}
  \REVIEW{Phys. Rev. Lett.}{104}{2010}{158001}.
    
\bibitem{Schockmel2013}
  \Name{Schockmel J., Mersch E., Vandewalle N., and Lumay G.}
  \REVIEW{Phys. Rev. E}{87}{2013}{062201}.
  
\bibitem{Ghazali2006}
  \Name{Ghazali A. and L\'{e}vy J.-C. S.}
  \REVIEW{Europhys. Lett.}{74(2)}{2006}{355}.
  
\bibitem{Dillmann2012}
  \Name{Dillman P., Maret G., and Keim P.}
  \REVIEW{J. Phys.: Condens. Matter}{24}{2012}{464118}.

\bibitem{Deutschlander2013}
  \Name{Deutschl\"{a}nder S. {\it et al.}}
  \REVIEW{Phys. Rev. Lett.}{111}{2013}{098301}. 
  
\bibitem{Boyer2009}
  \Name{Boyer F. and Falcon E.}
  \REVIEW{Phys. Rev. Lett.}{103}{2009}{144501}.
  
%20th
\bibitem{Coupier2005}
  \Name{Coupier G., Guthmann C., Noat Y., and Saint-Jean M.}
  \REVIEW{Phys. Rev. E}{71}{2005}{046105}.
  
\bibitem{Kohlstedt2005}
  \Name{Kohlstedt K. {\it et al.}}
  \REVIEW{Phys. Rev. Lett.}{95}{2005}{068001}.
  
\bibitem{SchefflerWolf2002}
  \Name{Scheffler T. and Wolf D.}
  \REVIEW{Gran. Mat.}{4}{2002}{103}.
  
\bibitem{MuellerLuding2011}
  \Name{M\"{u}ller M.-K. and Luding S.}
  \REVIEW{Math. Model. Nat. Phenom.}{6}{2011}{87}.
  
\bibitem{Shattuck}
  \Name{Shattuck M. D.}
  \REVIEW{"Particle tracking"}{\texttt{gibbs.engr.ccny.cuny.edu/technical/Tracking/ ChiTrack.php}}{accessed: 2013-04-18}{}.
   
\bibitem{CrockerWeeks}
  \Name{Crocker J. C. and Weeks E. R.}
  \REVIEW{"Particle tracking using idl"}{\texttt{www.physics.emory.edu/$\sim$weeks/idl/}}{accessed: 2013-04-18}{}. 
  
\bibitem{BlairDufresne}
  \Name{Blair D. and Dufresne E.}
  \REVIEW{"The matlab particle tracking code repository"}{\texttt{www.physics.georgetown.edu/matlab/}}{accessed: 2013-04-18}{}. 
  
\bibitem{Jackson}
  \Name{Jackson J.}
  \Book{Classical Electrodynamics, 3rd ed.}
  \Publ{Wiley, New York}
  \Year{1998}.
  
\bibitem{Prevost2004}
  \Name{Prevost A., Melby P., Egolf D. A., and Urbach J. S.}
  \REVIEW{Phys. Rev. E}{70}{2004}{050301}.
  
\bibitem{Falcon1999}
  \Name{Falcon E., Fauve S., and Laroche C.}
  \REVIEW{Eur. Phys. J. B}{9}{1999}{183}.
  
\bibitem{FalconWunenburger1999}
  \Name{Falcon E. {\it et al.}}
  \REVIEW{Phys. Rev. Lett.}{83(2)}{1999}{440}.
  
\bibitem{Strandburg1988}
\Name{Strandburg K. J.}
\REVIEW{Rev. Mod. Phys.}{60}{1988}{161}. 

\bibitem{Olafsen2005}
  \Name{Olafsen J. S. and Urbach J. S.}
  \REVIEW{Phys. Rev. Lett.}{95}{2005}{098002}.

\bibitem{Gradenigo2011}
  \Name{Gradenigo G., Sarracino A., Villamaina D., and Puglisi A.}
  \REVIEW{Europhys. Lett.}{96}{2011}{14004}. 
  
\bibitem{PuseyvanMegen1986}  
  \Name{Pusey P. N. and van Megen W.}
  \REVIEW{Nature}{320}{1986}{27}.
  
\bibitem{HanNature2008}
  \Name{Han Y. {\it et al.}}
  \REVIEW{Nature}{456}{2008}{898}.
  
\bibitem{ShokefPRL2009}
  \Name{Shokef Y. and Lubensky T. C.}
  \REVIEW{Phys. Rev. Lett.}{102}{2009}{048303}.

\bibitem{YaoPRL2013}
  \Name{Yao Z. and Olvera de la Cruz M.}
  \REVIEW{Phys. Rev. Lett.}{111}{2013}{115503}.  


  
\end{thebibliography}
\end{document}